\documentclass[10pt,twocolumn,letterpaper]{article}

\usepackage{cvpr}
\usepackage{times}
\usepackage{epsfig}
\usepackage{graphicx}
\usepackage{amsmath}
\usepackage{amssymb}

\newtheorem{definition}{Definition}
\usepackage[usenames]{color}
\usepackage{enumerate}
\usepackage{balance}

\newcommand{\figref}[1]{Figure~\ref{#1}}
\newcommand{\tabref}[1]{Table~\ref{#1}}
\newcommand{\secref}[1]{Section~\ref{#1}}
\newcommand{\equref}[1]{Equation~(\ref{#1})}

\usepackage{multirow}
\usepackage{tabularx}
\usepackage[boxed,ruled,vlined]{algorithm2e}
\usepackage[tight]{subfigure}
\newcommand{\hide}[1]{} %hide

\usepackage[breaklinks=true,bookmarks=false]{hyperref}

\cvprfinalcopy % *** Uncomment this line for the final submission

\def\cvprPaperID{****} % *** Enter the CVPR Paper ID here

\begin{document}

%%%%%%%%% TITLE
\title{Modeling Emotion Influence from Images in Social Networks}

\author{Xiaohui Wang, Jia Jia, Lianhong Cai, Jie Tang\\
Department of Computer Science and Technology\\
Tsinghua University\\
{\tt\small wangxh09@mails.tsinghua.edu.cn, jjia@tsinghua.edu.cn, } \\
{\tt\small clh-dcs@mail.tsinghua.edu.cn, jietang@tsinghua.edu.cn}
}

\maketitle
%\thispagestyle{empty}

%%%%%%%%% ABSTRACT
\begin{abstract}
Images become an important and prevalent way to express users' activities, opinions and emotions. In a social network, individual emotions may be influenced by others, in particular by close friends.
%We focus on emotions expressed by images and study an interesting problem of modeling emotion influence from images in social networks.
We focus on understanding how users embed emotions into the images they uploaded to the social websites and how social influence plays a role in changing users' emotions. We first verify the existence of emotion influence in the image networks, and then propose a probabilistic factor graph based emotion influence model to answer the questions of ``who influences whom''.
Employing a real network from Flickr as experimental data, we study the effectiveness of factors in the proposed model with in-depth data analysis. Our experiments also show that our model, by incorporating the emotion influence, can significantly improve the accuracy (+$5\%$) for predicting emotions from images. Finally, a case study is used as the anecdotal evidence to further demonstrate the effectiveness of the proposed model.
\end{abstract}

\section{Introduction}
\begin{figure*}[htb]
  \centering{\includegraphics[width=0.95\textwidth]{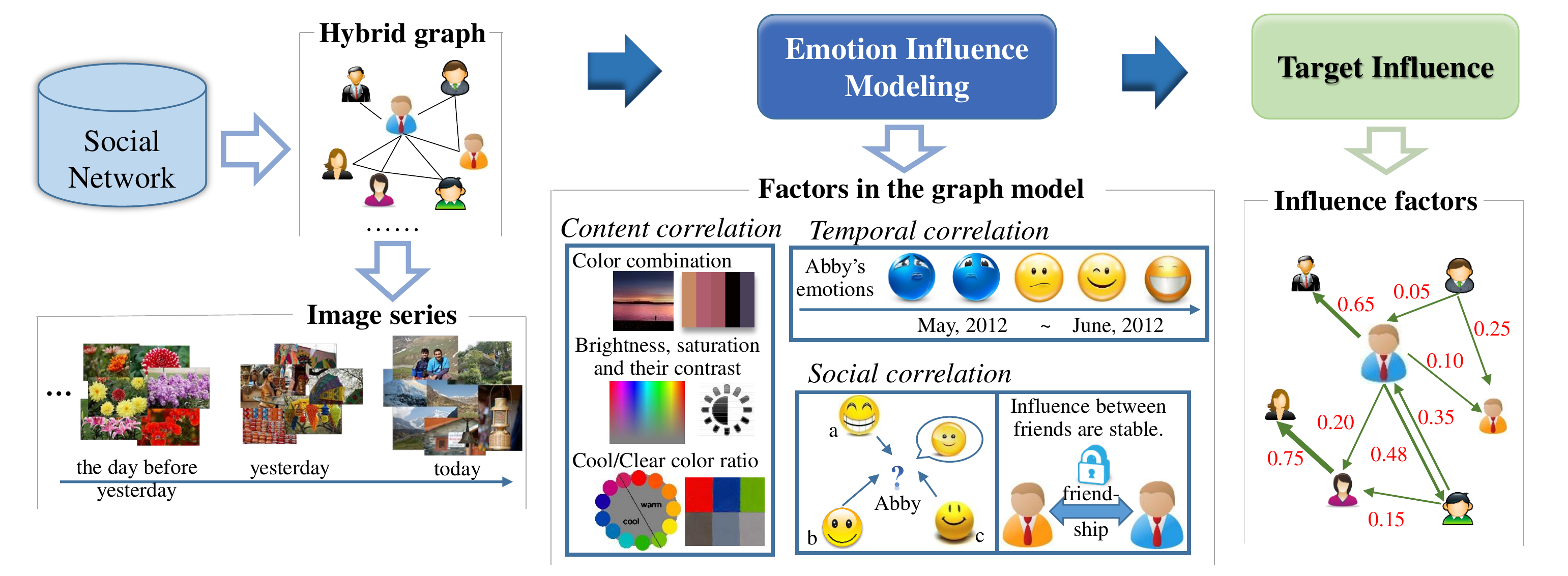}}
\caption{Illustration of modeling emotion influence from images in social networks.}
\label{fig:probIllus}
\end{figure*}

Emotion stimulates the mind 3,000 times quicker than rational thought~\cite{TangTAC}. One may make a quick decision simply because of a particular feeling (e.g., happy, sad, or angry). Recently, with the rapid proliferation of online social networks, such as Facebook, Twitter and Flickr, people start to express their daily feeling directly in the online networking space through texts and images. Understanding the substantial mechanism of emotion dynamics can give us insight into how
people's emotions influence their activities and how the emotions spread in the social networks.
%Therefore, there is a clear need for methods and techniques to analyze and quantify the dynamics of emotion influence in the social network.

In existing literature, several research studied the individual emotios in social networks from tweets, blogs and other personal attributes like locations and calling logs~\cite{happy2008,TangTAC}. However, these methods mainly considered the text information.
Therefore, they are difficult to be used to reveal the emotion dynamics in image-based social networks (e.g. Flickr, Instagram).
A public internet study on a Facebook dataset suggests that images drive event engagement 100 times faster (e.g., clicking ``like'' or adding comment)  than text updates.
Similar to the texts, images are also used to express individual emotions, but the expression is much more implicit.
%Different from text which can explicitly records (expresses) the emotional states, emotions are usually hidden in the images.
A simple example is that people often prefer to use warm colors like red or pink to express happiness and cold colors like blue or dark to express sadness.
Jia et al.~\cite{Jia12mm} tried to infer the emotions from Flickr images. They focus on the individual level, but ignore the emotion influence in social networks.
In social networks, individual emotions may be influenced by each other. The emotions hidden in their uploaded images also correlate with each other. How to leverage the correlation to help understand users' emotions is a challenging question.
Recently, the literature~\cite{machajdik2010:emotions} also studies the problem of inferring emotions from images using visual features. However, they do not consider the network information as well.

To clearly demonstrate the problem we are going to address in this paper, we give a general framework in~\figref{fig:probIllus}. The input of our study is an image-based social network comprised of users and user relations. Each user is associated with a set of images that she/he uploaded to the social network. The expected output is the learned emotion influence between users in the network.
The problem is non-trivial and has several challenges. Though a few literatures demonstrated the existence of influence in various social networks~\cite{Bakshy:12EC,Bond:12Nature,tang_social_2009}, it is still unclear whether such an influence also exists in the image networks.
A more challenging task is how to identify the emotion influence patterns, as users' emotions are usually affected by various complex and subtle factors.
Moreover, how to design a principled model to quantitatively describe the complex emotion influence among users and images?

In this paper, employing a networking data crawled from Flickr as our data source, we systematically study the problem of modeling emotion influence from images in social networks.
We first conduct a matched sampling test to justify the existence of influence in the image network. The influence test shows some interesting results: when one has more than two friends who uploaded ``happy'' images, the likelihood that the user also uploads ``happy'' images almost double the average probability.
We further propose a probabilistic factor graph model to formalize the emotion influence learning problem. In particular, the model considers the following factors:
(1) \textit{content correlation}: how images' visual features reflect users' emotions; (2) \textit{temporal correlation}: how a user's emotion is affected by her/his emotions in the recent past; (3) \textit{social correlation}: how users' emotions influence (and are influenced by) their friends' emotions.

Our experiments on the Flickr data demonstrate that the proposed model can achieve better performance with an average $5\%$ accuracy improvement than an alternative method using SVM. Based on the results, we investigate how to improve the performance by both visual and social attribute selection. Finally we demonstrate an interesting case study.

\section{Related work}
\label{sec:related}
%list influence paper£¬they use texts, we use imgs
\noindent{\textbf{Text-based social influence analysis.}} Previous research studied the influence in social networks based on users' actions~\cite{anagnostopoulos_influence2008}, opinions~\cite{la_fond_randomization_2010}, blogs and news articles~\cite{gomez-rodriguez_inferring_2012}.
%Goyal et al. proposed a method to learn the probabilities of influence between users in social networks~\cite{Goyal2010}.
However, they only use the text based information. In recently rapidly developing image based social networks, such as Flickr and Instagram, the text data or user actions are rather limited, and such text based methods can hardly work.
Modeling the influence problem on these image networks remains an interesting and challenging problem.
In this paper, we try to solve the problem of modeling the emotion influence in image based social networks.

\noindent{\textbf{Existence validation of emotion influence.}} Emotion influence is the basis and quite crucial for our image social network study.
Fortunately, some previous research in psychology and sociology has studied and confirmed the existence of emotion influence in social networks. Fowler et al. found that people's happiness depends on the happiness of others with whom they are connected~\cite{happy2008}. Some affective prediction methods have considered the emotional influence among users in social networks~\cite{Jia12mm,Wang12mm}. In this paper, we adopt Flickr data to validate the existence of emotion influence in image networks.

\noindent{\textbf{Affective image classification.}}
Emotional level image classification, also called affective image classification, is an important but hard problem.
This is because emotions are highly subjective, and difficult to quantitatively measure.
%V1
Previous research focused on two crucial aspects for classification accuracy: training data and models.
For the training data, the ground-truth emotions are usually manually labeled, therefore it is accurate but few in data number. Facing with the massive amounts of images in social networks, the above methods are powerless.
Recently, some work has predicted emotions from images in social networks with images' tags and comments~\cite{Jia12mm} or from users' actions~\cite{TangTAC}.
In this paper, we also adopt images' tags and comments to obtain the emotion labels as the ground truth.
For models, the commonly used methods are based on the machine learning such as Support Vector Machines (SVM)~\cite{classifSVM}, Naive Bayes~\cite{machajdik2010:emotions}, Random Forest~\cite{yanulevskaya_emotional_2008}, and probabilistic models recently~\cite{lafferty_conditional_2001,shin2010affective}.
Although these methods can achieve pretty good accuracies on affective image classification, they are difficult to incorporate different social influence factors.

\noindent{\textbf{Factor graph model in social network analysis.}}
The structure of the graphic model is similar as that of social networks, so it is commonly used for social analysis~\cite{TangTAC}.
%The graphic model has nature isomorphism with the social networks, making
A factor graph is a particular type of graph model that enables efficient computation of marginal distributions through the sum-product algorithm~\cite{kschischang_factor_2001}. Tang et al. analyzed social influence from texts using the factor graph model~\cite{tang_social_2009}.
In this paper, we adopt factor graph to build the emotion influence model, and the key problem is that what factors should be taken into account.

\section{Image-based Emotion Influence Study}
\label{sec:testInf}
\subsection{Data Collection}
\label{sec:obs_data}
We download a data set from Flickr\footnote{http://flickr.com}. The data set contains 4,725 users and their uploaded images (in total 1,254,640 images). Each user is associated with her/his shared images, contact list and personal information. Each image has timestamp, url, tags given by its owner and comments given by viewers.
%Each image has tags, descriptions and comments by the author and other viewers, as well as its timestamp, owner, url, etc.
%Each image is labeled with a number of tags, some of which directly reflect emotional states.

\hide{We use the traditional lexicon-based approach to assign the emotion of each image. To avoid ambiguity, we use WordNet~\cite{Wordnet} and HowNet~\cite{Hownet} dictionaries to obtain more than $200$ synonyms for each of the six emotional categories, and manually verify them. For each image, we count the occurrences of each emotion synonyms in its tags and comments, and choose the most frequency one (if exists).
Finally, 50,210 images are labeled by this method and adopted in our data observation.}

In our problem, the emotions are defined as six basic categories according to Ekman's theory~\cite{ekman1992}, \textit{happiness, surprise, anger, disgust, fear} and \textit{sadness}. Taking~\cite{Jia12mm} as a reference, we use WordNet~\cite{Wordnet} and HowNet~\cite{Hownet} dictionaries to obtain more than $200$ synonyms for each emotional category, and manually verify them. For each image, we count the occurrences of each emotion synonyms in its tags and comments, and select the most frequency one (if exists) as the ground truth. Finally, 50,210 images are labeled by this method.

%\begin{table}
%\label{tab:synonym}
%\centering
%\caption{Examples and number of synonyms per emotional category.}
%\begin{tabular}{|m{1.3cm}|m{0.4cm}|m{5.7cm}|} \hline
%Category & \# & Synonym examples \\ \hline
%Happiness & 177 & pleasure, charming, joyous, brilliant, glad, delight, sweet, enjoyable, satisfying\\ \hline
%Surprise & 39 & shock, startle, surprising, amazed, astonish, astonied\\ \hline
%Anger & 442 & annoyance, fury, ire, irritation, rage, wrath, storming, aggravate\\ \hline
%Disgust & 241 & nauseate, offend, repel, revolt, sicken, nasty\\ \hline
%Fear & 102 & horror, terror, frightful, dread, scare, dismay, direful\\  \hline
%Sadness & 432 & depressed, unhappy, miserable, discourage, moody, cried, displeasing\\ \hline
%\end{tabular}
%\end{table}

\subsection{Sampling Test}
\label{sec:Sampling Test}
We first need to validate the existence of emotion influence in image networks. As a premise, we assume that uploaded images can reflect their owners' emotions. The basic validation method is the sampling test~\cite{JingICJAI13}. The users are divided into two groups: the friend-related group $G_R$ and the friend-independent group $G_I$. Taking the ``happy'' emotion as an example,
a user is said to be ``happy'' if most of her/his uploaded images are labeled ``happy'' at a specified time.
$G_R$ contains users who uploaded images at time $t$ and has one or more friends with the ``happy'' emotion at time $t-\Delta t$, while $G_I$ contains users who uploaded images at time $t$ and has no friend with the ``happy'' emotion at time $t-\Delta t$. Finally, we compare the ``happy'' ratio of users at time $t$. The ``happy'' ratio is defined as the ratio of ``happy'' users in each group, e.g. $Ratio=\frac{\#happy\_users}{\#all\_users}$.

\noindent\textbf{Experimental setting.} For each group, $50$ users are randomly chosen from our data set. We set $\Delta t={1,2,3,4}$ weeks and $t$ is randomly chosen from March, $2006$ to June, $2012$ without overlap. We repeat the experiment $10$ times to calculate the average ratio. Intuitively, the more friends have the same emotion, the greater is their influence. To investigate the effect of the number of friends, we further divide the friend-related group $G_R$ into two subgroups: subgroup with one or two friends and subgroup with at least three friends (with the ``happy'' emotion at time $t-\Delta t$).

\begin{figure}[t]
  \centering
    \includegraphics[width=8.5cm]{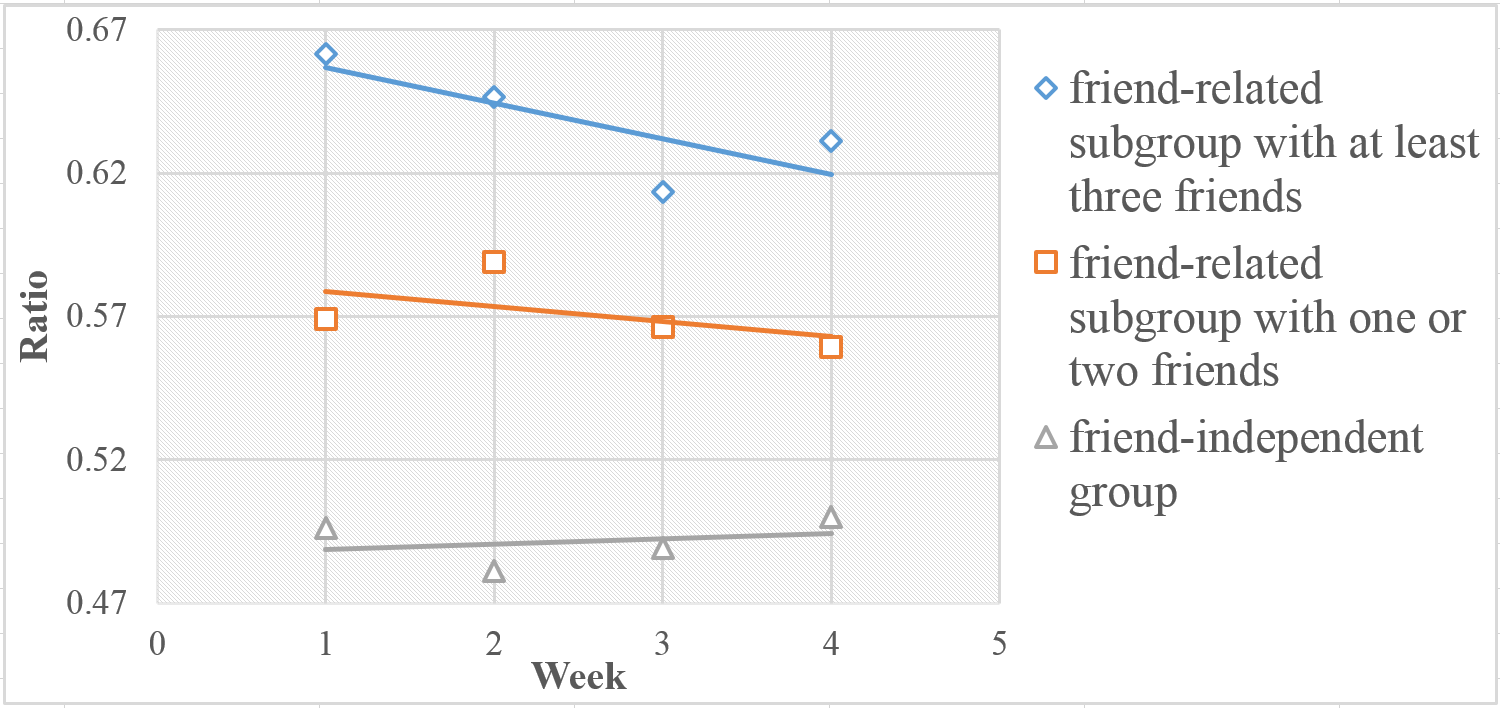}
\caption{The result of sampling test for image-based emotion influence.}
\label{fig:PNAStest}
\end{figure}

\noindent\textbf{Results.} \figref{fig:PNAStest} shows the results. The average ratio of the friend-related group $G_R$ is significantly higher than that of the friend-independent group $G_I$, which confirms the existence of friend influence through images. The results also indicate more friends with the same emotion usually have bigger impact. And the downtrend shows that the influence becomes weakened as time passed, which will be considered in our model.

\section{Problem Formulation}
\label{sec:prob}
%Our goal is to derive the emotional states from users' uploaded images and analyze the emotion influence in social networks.
In this section, we present the problem formulation of modeling the emotion influence from images in social networks.
The social network can be defined as a graph $G=(V, E)$, where $V$ is the set of $|V|=N$ users, $E \subset V \times V$ is the set of relationships among users, e.g. friendships. Notation $e_{ij} \in E$  indicate user $v_i$ and $v_j$ are friends in the social network.
To incorporate the temporal factors, we also divide the continuous time to time slices. So here ``at time $t$'' refers to time slice $t$.
A user $v_i$ uploaded a set of images at time $t$, denoted as $\mathbf{X}_i^t$. $\mathbf{x}_{ij}^t$ is an element of $\mathbf{X}_i^t$ indicating the $j$th image uploaded by user $v_i$ at time $t$, which is instantiated as its visual features.~\tabref{table:nots} summarizes the notations used throughout this paper.

%In our problem, the emotion space is defined as six most common emotional states, \textit{happiness, surprise, anger, disgust, fear} and \textit{sadness}~\cite{ekman1992}.
Each image is associated with one type of six emotions and our first task is to classify the emotion category of each image.
Without loss of generality, we treat it as a binary classification problem, and run the training/inference for each emotion separately. Let $y_{ij}^t,y_i^t \in \{-1,1\}$ be binary variables indicating whether an image $\mathbf{x}_{ij}^t$ or a user $v_i$ at time $t$ has a specified emotion.
When all models suggest a negative value for an image/user, it means that the image/user has a neutral emotion. When multiple models suggest positive values, we select the emotion with the highest probability (Cf. \secref{sec:method} for the definition of the probability).

\begin{definition}
\textbf{Emotion influence.} Social influence from user $i$ to user $j$ at time $t$ is denoted as $\mu_{ij}^t$. In our implementation, we treat is as a binary variable $\{0,1\}$, and use the inferred possibility as the output weight.
\end{definition}

\begin{definition}
\textbf{Time-varying social network.} The time-varying social network is denoted as $G=(V, \{E^t\}, \{\mathbf{X}_i^t\})$, where $V$ is the set of users, \hide{$E$ is the set of relations with time information,} $e_{ij}^t\in E^t$ denotes user $v_i$ and user $v_j$ are friends at time $t$, $\mathbf{X}_i^t$ denotes images uploaded by user $v_i$ at $t$. %Note that we only consider variables related to a user after his registration.
\end{definition}

\noindent\textbf{Learning task:} Given a time-varying social network $G$, the target is to find a function for predicting emotions of all the unlabeled images, users' emotions and influence in different time slices:
\begin{equation}
    f: (G, labeled~data) \rightarrow (\{y_i^t\}, \{y_{ij}^t\}, \{\mu_{ij}^t\})
\end{equation}

Learning the function depends on multiple factors, e.g., emotions of images, users' emotions and users' correlations. The challenge is how to design a unified model that could incorporate all these factors together.
%Besides, the scale of the social network is very large.

\begin{table}
\centering
\caption{Notations}
\label{table:nots}
\footnotesize
\begin{tabular}{|c|m{6.8cm}|} \hline
Symbol&description\\ \hline
$V$ & the set of users in the social network\\ \hline
$E^t$ & the set of edges at time $t$\\ \hline
$N$ & number of users \\ \hline
$\mathcal{T}$ & the collections of all the successive time slices, namely 1,2,3...\\ \hline
$\mathbf{x}_{ij}^t$ & the vector of visual features of the $j$th image uploaded by user $v_i$ at time $t$ \\ \hline
$\mathbf{X}_i^t$ & the set of images uploaded by user $v_i$ at time $t$\\ \hline
$NB^t(v_i)$ & the set of $v_i$'s friends at time $t$\\ \hline
$y_{ij}^t$ & binary variable indicating whether image $\mathbf{x}_{ij}^t$ has a specified emotion\\ \hline
$y_{i}^t$ & binary variable indicating the emotion of user $v_i$ at time $t$\\ \hline
$\mu_{ij}^t$& binary variable indicating the influence of user $v_i$ on $v_j$ at time $t$
\\
\hline
\end{tabular}
\end{table}

\section{Emotion Influence Model}
\label{sec:method}
We propose a factor graph model to infer emotion influence from images in social networks (\figref{fig:probIllus}).
We consider the following aspects:
(1) \textit{content}: a user's emotion is induced by visual features of their uploaded images;
(2) \textit{time}: the user's current emotion has correlations with her/his emotion in the recent past;
(3) \textit{influence}: the user's emotion may be largely affected by his friends, and the influence relationship does not change frequently within a short time.
By leveraging these aspects, we formulate the emotion influence model as a dynamic factor graph model.
\begin{figure}[htb]
  \centering{\includegraphics[width=0.95\linewidth]{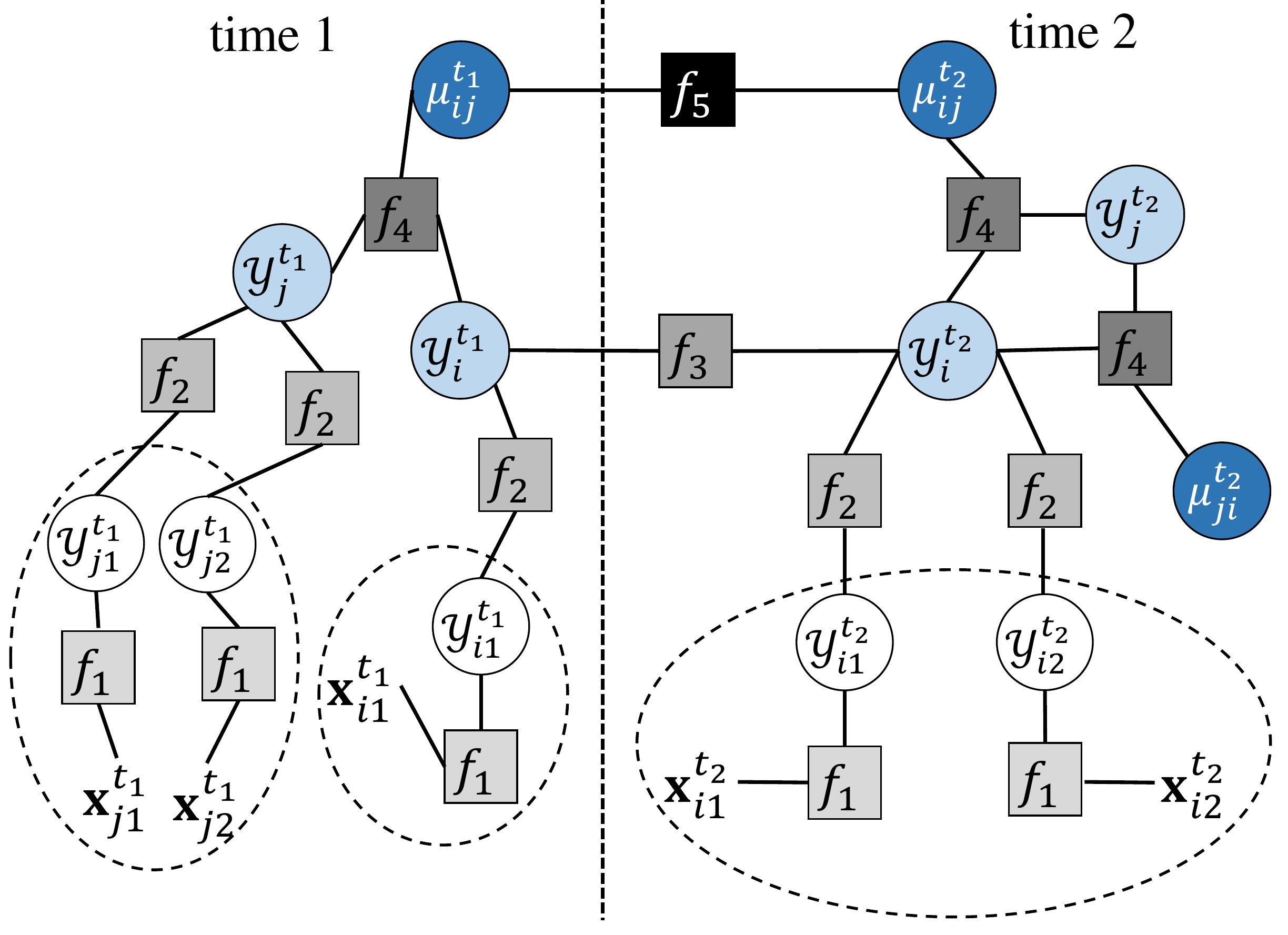}}
\caption{Graphical representation of the emotion influence model.}
\label{fig:model}
\end{figure}
\subsection{The Predictive Model}
In the model, the above aspects are quantized as different types factor functions, as illustrated in \figref{fig:model}:
\begin{itemize}
\item \textbf{Content function.} It contains two parts, image induction factor $f_1(y_{ij}^t, y_{i}^t)$, denoting the emotions of uploaded images reflect their owners' emotion, and visual feature factor $f_2(\mathbf{x}_{ij}^t, y_{ij}^t)$, representing the correlation between image emotion $y_{ij}^t$ and its corresponding visual feature $\mathbf{x}_{ij}^t$.
\item \textbf{Temporal function.} $f_3(y_i^{t'},y_i^t), t'<t$. It represents the influence of user $v_i$'s emotions in the recent past $t'$ on her/his current emotion at time $t$.
\item \textbf{Social function.} $f_4(y_i^t,y_j^t, \mu_{ij}^t)$. It encodes the dependence of user $v_i,v_j$'s emotion at time $t$, and their influence variable $\mu_{ij}^t$. To better constrain the influence, we assume that the influence between two users are stable over time, encoded by $f_5(\mu_{ij}^{t'},\mu_{ij}^{t}), t'<t$.
\end{itemize}

%These functions can be defined as many different types. In this paper, we give the general definitions.
The basic assumption here is that a user's emotion at some time can be reflected by ``emotions'' of the images she/he uploaded around that time. Thus we could define the following factor function:
\begin{equation}
f_1(y_{ij}^t, y_i^t)=\exp\{-\beta_i |y_{i}^t-y_{ij}^t|\}
\end{equation}

Visual feature factor $f_2$ is instantiated as exponential-linear function:
\begin{equation}
f_2(\mathbf{x}_{ij}^t, y_{ij}^t)=\exp\{\alpha^T \cdot \mathbf{x}_{ij}^t y_{ij}^t\}
\end{equation}
where $y_{ij}^t \in \{-1,1\}$ is the binary emotion indicator. $\mathbf{x}_{ij}^t$ is the visual feature, $\alpha$ is the parameter vector for all the images. And $f_2$ can be thought of a weak predictor, working directly on images.
We adopt the interpretable aesthetic features~\cite{Wang13ICIP} as our visual features, as summarized in \tabref{table:features}. It contains several color features, including the well known five-color combination~\cite{kobayashiFiveColor1995}, which has direct impact on image emotions~\cite{itten1974art}.

\begin{table*}
  \centering
  \caption{Summary of visual features. The column `\#' represents the dimension of each feature.} \label{table:features}
  \footnotesize
  %\begin{tabular}{|m{2.2cm}|m{0.15cm}|m{5.0cm}|} \hline
  \begin{tabular}{|m{2.8cm}|m{0.2cm}|m{13.2cm}|} \hline
     Name & \# & Short description \\\hline
     Five-color combinations & 15 & five dominant colors in the HSV color space \\ \hline
     Brightness, saturation and their contrast & 4 & mean brightness, mean saturation and their average contrast \\\hline
     Cool color ratio & 1 & ratio of color colors. Colors can be divided into cool colors with hue ([0,360]) in the HSV space between 30 and 110 and warm colors. \\\hline
     Clear color ratio & 1 & ratio of clear colors with brightness ([0,1]) greater than 0.7. The colors whose brightness less than 0.7 are dull colors. \\\hline
  \end{tabular}
\end{table*}

%时间是可以累积的
Usually, a user's emotion do not change rapidly and her/his current emotion is highly dependent on her/his emotions in the recent past. So the temporal correlation factor function is used to model this phenomenon:
\begin{equation}
f_3(y_i^{t'},y_i^t)=\exp\{-\xi_i\cdot e^{-\delta_i\cdot|t'-t|}|y_i^t-y_i^{t'}|\}
\end{equation}
%where $t' \in T$ is the past period and within the influencing and effective period $T$. We accumulate effects of the temporal correlation, that is $t'$ is taken different values.
%and the effects are reduced as the time interval are get longer.
where $e^{-\delta_i\cdot|t'-t|}$ indicates the correlation decline effects over time,and $\xi_i,\delta_i$ are the per-user weight parameters.

As obtained in data observation in~\secref{sec:test}, the emotion can be affected by his friends. And the social influence factor function is defined to model this effect:
\begin{equation}
f_4(y_i^{t},y_j^t,\mu_{ij}^t)=\exp\{-\lambda_{ij}|1-\mu_{ij}^t-|y_i^t-y_j^{t}||)\}
\end{equation}
where $\mu_{ij}^t\in\{0,1\}$ is a binary variable, representing at time t, whether user $v_i$ has influence on user $v_j$. $\lambda_{ij}$ is the per user pair weight parameter.
%Similarly as in $f_3$, The social influence is accumulated over time.

For stable influence factor function, we assume that the social influence is stable, which means if a friend has a strong influence on you before, the impact is likely to be also strong afterwards. This constraint is defined as follows:
\begin{equation}
f_5(\mu_{ij}^{t'},\mu_{ij}^{t})=\exp\{-\eta_{ij} e^{-\tau_{ij}\cdot|t'-t|} |\mu_{ij}^t-\mu_{ij}^{t'}|\}
\end{equation}
where $e^{-\tau_{ij}\cdot|t'-t|}$ indicates the decay effect over time and $\tau_{ij},\eta_{ij}$ are two weight parameters.

%The objective function is defined as the joint probability of the above factor functions:

Given the definitions of the above factor functions, we can define the following log-likelihood objective function:
\small
\begin{equation}
\label{eq:objective}
\begin{split}
\mathcal{L}=& \sum_{v_i\in V,t\in \mathcal{T}} \sum_{\mathbf{x}_{ij}^t \in \mathbf{X}_i^t} -\beta_i |y_i^t-y_{ij}^t| + \sum_{v_i\in V,t\in \mathcal{T}} \sum_{\mathbf{x}_{ij}^t \in \mathbf{X}_i^t} \alpha^T \cdot \mathbf{x}_{ij}^t y_{ij}^t \\
&+ \sum_{v_i\in V,t,t'\in \mathcal{T}}-\xi_i\cdot e^{-\delta_i\cdot|t'-t|}|y_i^t-y_i^{t'}| \\
&+ \sum_{v_i\in V,t\in \mathcal{T}} \sum_{e_{ij}^t \in E^t} -\lambda_{ij}|1-\mu_{ij}^t-|y_i^t-y_j^{t}|| \\
&+ \sum_{v_i\in V,t,t'\in \mathcal{T}} \sum_{e_{ij}^{t'} \in E^{t'}, e_{ij}^t \in E^t}
%NB^t(v_i)\bigcap NB^{t'}(v_i)}
-\eta_{ij} e^{-\tau_{ij}\cdot|t'-t|} |\mu_{ij}^t-\mu_{ij}^{t'}| -\log Z
\end{split}
\end{equation}
\normalsize
where $Z$ is a normalization factor. % and $v_i\sim $$
%$NB(i)$ denotes the set of friends of user $v_i$.
%; $T$ is a valid time interval, and $t'\in T$ denotes the influence accumulation over the time interval.

%$\alpha_{ij}$ is denoted as their familiarity, eg. for two friends, it can be defined as their common friends (needs to be normalized); $\beta_{ij}^t$ is defined as their interaction factor at period $t$, eg. their common posts with the same emotions.

\subsection{Model Learning and Inference}
\begin{algorithm}
\label{mainalgo}
\SetAlgoLined
\caption{The model learning/prediction algorithm}
\KwIn{Graph connection, image features, image labels}
\KwOut{Parameters: $\alpha$, $\{\beta_i\}$, $\{\xi_i\}$, $\{\delta_i\}$, $\{\lambda_i\}$,  $\{\eta_i\}$, $\{\tau_i\}$,
Variables: $\{y_i^t\}$, $\{y_{ij}^t\}$, $\{\mu_{ij}^t\}$}
Initialize $\alpha=$predictor vector by linear SVM, $\beta_i=0.6$, $\xi_i=0.5$, $\delta_i=1$, $\lambda_i=0.1$, $\eta_i=0.5$, $\tau_i=1$\\
\Repeat{convergence}
{
    1. Assume $\alpha$, $\{\beta_i\}$, $\{\xi_i\}$, $\{\delta_i\}$, $\{\lambda_i\}$,  $\{\eta_i\}$, $\{\tau_i\}$ fixed, run max-product belief propagation to infer variables $y_i^t$, $y_{ij}^t$, $\mu_{ij}^t$\;

    2. Fix variables $y_i^t$, $y_{ij}^t$, $\mu_{ij}^t$, parameters $\{\delta_i\}$, $\{\tau_i\}$, use iterative gradient ascend to find parameters $\alpha$, $\{\beta_i\}$, $\{\xi_i\}$, $\{\lambda_i\}$, $\{\eta_i\}$;

    3. Fix variables $y_i^t$, $y_{ij}^t$, $\mu_{ij}^t$, parameters $\alpha$, $\{\beta_i\}$, $\{\xi_i\}$, $\{\lambda_i\}$,  $\{\eta_i\}$, use gradient ascend to find parameters
$\{\delta_i\}$, $\{\tau_i\}$
}
\end{algorithm}

Due to the flexibility of the model, there are quite a lot of free parameters in the formulation, including $\alpha$, $\{\beta_i\}$, $\{\xi_i\}$, $\{\delta_i\}$, $\{\lambda_{ij}\}$, $\{\mu_{ij}\}$, $\{\tau_{ij}\}$, where $\xi_i$, $\delta_i$ iterate over users, and $\lambda_{ij}$, $\mu_{ij}$, $\tau_{ij}$ iterate over user pairs. To avoid possible over-fittings, we transform these user pair variables to one user variables. That is, we assume $\lambda_{ij}=\lambda_{i}$, $\eta_{ij}=\eta_i$, $\tau_{ij}=\tau_i$, $\forall j\in V$. We also have the binary variables to infer, including $y_i^t$, $y_{ij}^t$ and $\mu_{ij}^t$. And our goal is to maximize the objective function \equref{eq:objective}.

Basically, there are two steps, namely ``learn'' and ``predict''. That is, based on the emotion influence model and labeled data, we estimate a parameter configuration. And then these parameters are used to infer the variables. However, it is pretty hard to find an approximating well labeled sub-network for us to train. That is, even for the same user at the same time, some images are labeled and some are not. Here we combine the two processes into one optimization framework. Algorithm \ref{mainalgo} gives the details.

Generally, we use a fix and update iterative approach to predict the variables and learn the parameters. Note that some variables ($y_{ij}^t$) are already labeled, and treated as constants during optimization. In each iteration, first, a traditional max-product belief propagation is used to infer the variable according to current parameters.
During the second step, we assume parameters $\{\delta_i\}$, $\{\tau_i\}$ are fixed, it is an MRF parameter estimating procedure. In this scenario, \equref{eq:objective} can also be written as:
\begin{equation}
p(\mathbf{q}|\mathbf{\theta})=\frac{1}{Z_{\mathbf{\theta}}} \exp(\mathbf{\theta}^T \mathbf{\phi}(\mathbf{q}))
\end{equation}
where $\mathbf{q}$ is the variable vector, and let $\mathbf{q_0}$ be its current value.

The log-likelihood objective function is:
\begin{equation}
\mathcal{O}(\mathbf{\theta})=\log p(\mathbf{q_0}|\mathbf{\theta})=\mathbf{\theta}^T \mathbf{\phi}(\mathbf{q_0})-\log Z_{\mathbf{\theta}}
\end{equation}

%To use the gradient-based optimizer, the gradient is
%\begin{equation}
%\frac{\partial \mathcal{O}}{\partial \mathbf{\theta}}=\mathbf{\phi}(\mathbf{q_0})-
%\frac{\partial}{\partial \mathbf{\theta}}\log Z_{\mathbf{\theta}}
%\end{equation}
%where $Z_{\mathbf{\theta}}$ is the sum of all possible configurations of $\mathbf{q}$. And the gradient of $Z_{\mathbf{\theta}}$ is given by
%\begin{equation}
%\frac{\partial \log Z_{\mathbf{\theta}}}{\partial \mathbf{\theta}}=
%\sum_{\mathbf{q}}\phi(\mathbf{q})p(\mathbf{q}|\mathbf{\theta})=
%\mathbb{E}[\phi(\mathbf{q})|\mathbf{\theta}]
%\end{equation}
%Hence the gradient of the log-likelihood objective function is
%\begin{equation}
%\frac{\partial \mathcal{O}}{\partial \mathbf{\theta}}=\phi(\mathbf{q}_0) - \mathbb{E}[\phi(\mathbf{q})]
%\end{equation}
Its gradient is
\begin{equation}
\frac{\partial \mathcal{O}}{\partial \mathbf{\theta}}=\phi(\mathbf{q}_0) - \mathbb{E}[\phi(\mathbf{q})]
\end{equation}
And then a standard gradient method can be used to estimate the parameters.

Note that a similar gradient calculation can be applied to the third step. But it may loose the convexity as step 2 has. Fortunately, the step 3 parameters only controls correlation decaying speed over time, and are usually common among users. Even fixed parameters work well. So the non-convexity does not affect the effectiveness of our method.

%First the parameters are initialized randomly. Then we update the parameters by $\mathbf{\theta}_{new}=\mathbf{\theta}_{old}+\gamma \frac{\partial \mathcal{O}}{\partial \mathbf{\theta}}$, where $\gamma$ is the rate of convergence. This process iterates times until convergence.
%
%Given the graph and the learned parameters $\mathbf{\theta}$, the emotion of images can be iinferred as follows:
%\begin{equation}
%\mathbf{Y}=\arg \max_{\mathbf{Y}} p(\mathbf{Y}|\mathbf{X},\mathbf{\mu},\mathbf{\theta})
%\end{equation}
%
%Another inference task is to obtain the emotional influence between friends in the graph, given by:
%\begin{equation}
%\mathbf{\mu}=\arg \max_{\mathbf{\mu}} p(\mathbf{\mu}|\mathbf{X},\mathbf{Y},\mathbf{\theta})
%\end{equation}

\section{Data Observation}
\label{sec:test}
In this section, we evaluate the rationality of the model factors by statistical data observations.

\subsection{Visual Features}
%sad- more blue color; happy-warm colors£¬sad-cool colors
Factor $f_2$ is determined by visual features of images.
%We use statistics to confirm the strong relations between the visual features and image emotions.
% show difference
%\textbf{Color Feature Distributions.} We calculate color feature distributions for each emotion and found that for different emotions, the distributions share some common characters, but also have significant differences. \figref{fig:featureDif} shows an example illustrating the hue and saturation distributions under different emotions. In \figref{fig:featureDif} (left), the sadness distribution (black) is significantly above the happiness (white) near the blue segments, conforming to human intuition. Intuitively, the saturation expresses the degree of vividness, and warm colors with high saturation usually make people happy, while cool colors with low saturation can stimulate negative feelings. In \figref{fig:featureDif} (right), the saturation curve of happiness (yellow) is displaced to right, compared with other negative emotions, consistent with the above point.
%We calculate visual feature distributions for each emotion. From \figref{fig:featureDif} we can find that for different emotions, the distributions have some differences, but share more common characters.
We test whether there exist explicit correlations between the visual features and emotions by Canonical correlation analysis (CCA). CCA is a statistical approach suitable for multidimensional data~\cite{hardoon_canonical_2004}, which can explicitly show that how related two random vectors are, to the maximal possible extent compared with simple correlation methods like Pearson analysis.
%Visual features (such as color combinations) are believed to have strong relationship with the image-scale space.
To comply with the requirements of CCA, emotions should be quantitatively described. Here we adopt the image-scale space, composed of two dimensions \textit{warm-cool} and
\textit{hard-soft}~\cite{kobayashiFiveColor1995}, to represent emotions. Each labeled image in our data set is assigned by the image scale according to its tags and comments using the method~\cite{Wang12VisualComputer}.
We have two sets of variables, $X=\{x_1,x_2,..,x_{21}\}$ representing visual features in \tabref{table:features} and $Y=\{wc,hs\}$ indicating image scale. CCA will find linear combinations of the $x$'s and the $y$'s that have maximum correlation with each other.
The final results on our data set show the maximum correlation coefficient $0.74$, indicating that there exists considerable correlation between the visual features and emotions.
%\begin{figure}[t]
%  \centering
%    \includegraphics[width=3.5cm]{image/emo_top_rgb_all5color.png}
%    \includegraphics[width=4.3cm]{image/emo_rgb_nogrey_saturation.png}
%\caption{Examples of visual feature distributions.}
%\label{fig:featureDif}
%\end{figure}
\subsection{Temporal Correlation}
\begin{figure}[htb]
  \subfigure[Temporal correlation]{
    \label{fig:mini:temporalObs:a}
    \begin{minipage}[b]{0.5\linewidth}
      \centering
      \includegraphics[width=1.0\linewidth]{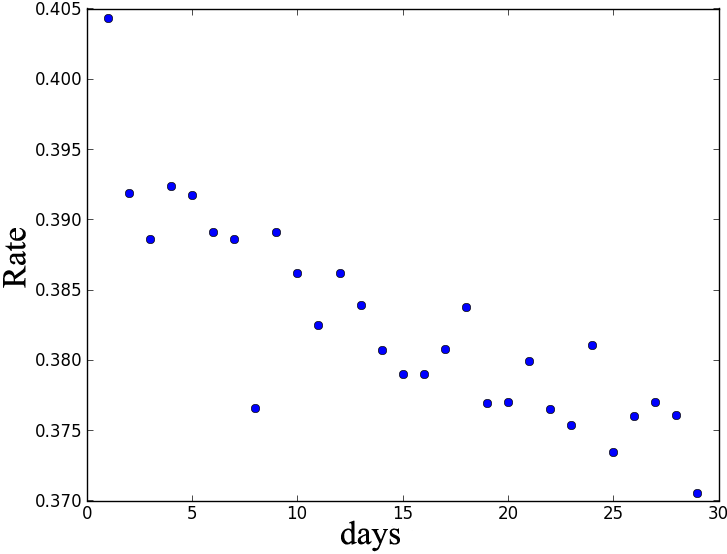}
    \end{minipage}}%
  \subfigure[Social influence]{
    \label{fig:mini:temporalObs:b}
    \begin{minipage}[b]{0.5\linewidth}
      \centering
      \includegraphics[width=1.0\linewidth]{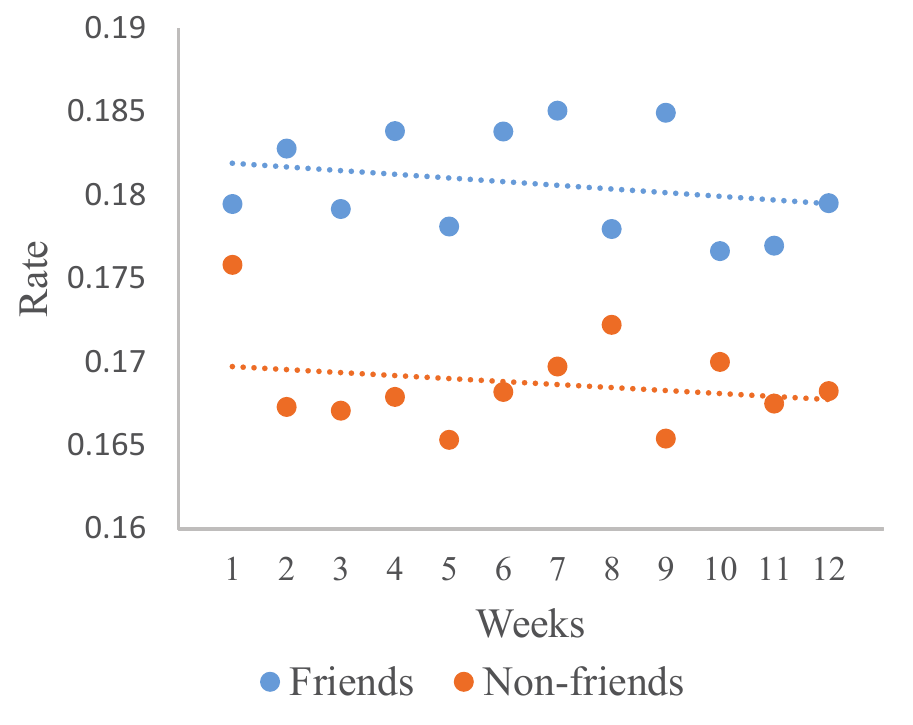}
    \end{minipage}}
  \caption{Data observation results.}
  \label{fig:mini:temporalObs}
\end{figure}

To better illustrate the temporal factor $f_3$, we track the users' emotion over time.
\figref{fig:mini:temporalObs:a} shows the temporal correlation rate $Rate_T$ of randomly selected $2500$ users ($n=2500$) during a month ($\delta=1,2,..,29$). $Rate_T$ is the average rate of users with the same emotion between time $t$ and $t+\delta$, defined as:

\begin{equation}
\label{rateT}
    Rate_T=\frac{1}{n}\sum_{i=1}^{n}\frac{\sum{\#Same\_emotion(v_i^t, v_i^{t+\delta})}}{\#\{t\}}
\end{equation}
%As in \secref{sec:Sampling Test}, a user's emotion is determined by his uploaded images.
$\#\{t\}$ is number of tested $t$ and $t+\delta$ pairs for user $v_i$ (user $v_i$ uploaded images at the two time points).
The overall tendency is downward, showing that the emotion similarity drops for longer time intervals. This confirms the correlation between a user's current emotion and her/his emotions in the recent past.

\subsection{Social Correlation}
To better examine the influence factors $f_4$ and $f_5$, we generalize the statistical view of the social correlation in \secref{sec:Sampling Test}, by counting general user pairs and test whether they have the same emotions. This test further tell us whether one's emotions could influence her/his friends.
%We compare the similarity between a user's emotion in a period and his/her friends afterward.
We randomly choose $n=1000$ users and their uploaded images. \figref{fig:mini:temporalObs:b} shows the similarity rate $Rate_I$ for friends and non-friends, defined as:
\begin{equation}
\label{rate2}
    Rate_I=\frac{1}{n}\sum_{i=1}^n\frac{\#Same\_emotion(v_i^t, NB(v_i)^{t+\delta})}{\#NB(v_i)}
\end{equation}
\noindent where $Same\_emotion(v_i^t,NB(v_i)^{t+\delta})$ is a function counting the number of users of group $NB(v_i)$ at time $t+\delta$ with the same emotion as user $v_i$ at time $t$. Note that we do not require members of $NB(v_i)$ to have an emotion at $t+\delta$. $\delta$ is the time interval, ranging from 1 to 12 weeks.
For friends test, $NB(v_i)$ is defined as friends of $v_i$, and for non-friends test, $NB(v_i)$ is chosen randomly.
%We randomly choose $\#NB(v_i)$ nodes except $v_i$'s connected nodes and recalculate the rate. If the social influence does not exist, the two rates almost the same.
The result shows the influence rate of friends is greater than that of non-friends, indicating significant emotion influence among friends.

\section{Experiments and Results}
\label{sec:exp}
%   1. accuracy,与svm比较
%   2. iteration number--fig8曲线
%   3. factor function contribution analysis: analyze feature contribution, eg. t=1,3,7 day，去掉一个function分析准确率的变换
%   4. case study: svm is wrong, our method is right. why, eg. because the graph relationship
% case study:影响的大小通过交互的条数来验证，画一个分布图，边标注条数和算得的影响。
%圣诞节前后的情感比例（图片和人两部分）
In this section, we evaluate the emotion prediction accuracy of images using the proposed model and then analyze how social factors help improve the inferring performance. Finally, we give a qualitative case study to further demonstrate the effectiveness of the method. The dataset is described in \secref{sec:obs_data}. We use 50,210 labeled images and their corresponding users to train and test the proposed model. All the images and users are adopted to infer the emotion influence.

%\subsection{Experimental Setup}
%The data set is extracted from Flickr. There are 4,725 users with 1,254,640 images, in which 50,210 images have emotion labels (the same as \secref{sec:obs_data}). We use 50,210 labeled images and their corresponding users to train and test the proposed model.
%The number of images per emotional category is shown in~\tabref{tab:imgnum}.
%The total 1.25 million images and all the users are used to infer the emotion influence.
%\begin{table}
%\centering
%\caption{Number of images per emotional category.}
%\begin{tabular}{|c|c|c|c|c|c|}\hline
%\centering
%Happiness & Surprise & Anger & Disgust & Fear & Sadness\\ \hline
%30727 & 4124 & 1578 & 330 & 10026 & 3425\\ \hline
%\end{tabular}
%\label{tab:imgnum}
%\end{table}

\subsection{Model Performance}
We compare the proposed model with an alternative method using Support Vector Machine (SVM) for emotion prediction of images. %Similar to the logistic regression model, SVM uses attributes associated with each edge as features to train a predictive model and then employs the model to predict edges' labels in the test data set.
SVM directly uses the visual features to train a predictive model.
We use the emotion labels of 50,210 images as the ground truth of emotion prediction. We evaluate the performance by accuracy and F1-Measure. Accuracy is the proportion of true results (both true positives and true negatives), while F1-Measure is calculated by Precision and Recall.

For all the six emotional categories, the average accuracy of the proposed model achieves 75.00\%, while the average F1-Measure is 40.48\%. Considering that our experimental data contains more than 50 thousand images from online social networks, the results are quite encouraging compared with the traditional work of predicting image emotion. \tabref{tab:factorContr} shows the comparison performance results of our model and SVM. Using the same features, we find that our model can achieve an average 5\% improvement on Accuracy and average 3\% improvement on F1-Measure than SVM, which indicates our model can better describe the social structure in our formulated problem.
%\begin{figure*}[htb]
%  \centering
%  \subfigure[Accuracy]{\includegraphics[width=0.45\linewidth]{image/acc_svm.pdf}}
%  \subfigure[F1-Measure]{\includegraphics[width=0.45\linewidth]{image/f1_svm.pdf}}
%\caption{Performance comparison between our model and SVM.}
%\label{fig:comp}
%\end{figure*}

\subsection{Factor Contribution Analysis}
In the predictive model, we incorporate both visual factors and social factors. Here we investigate how social factors help improving the performance of inferring image emotions. We have defined three social factor functions: temporal correlation ($f_3$), social influence ($f_4$) and stable influence ($f_5$). We test the contribution for each factor function by removing it from the model in turn and compare the prediction performance.

\tabref{tab:factorContr} shows the results of social factor contribution analysis. For most emotional categories, we can see the descending on both Accuracy and F1-Measure when removing each of the three social factors. Previous researches have revealed that social factors play an important role in text-based prediction task in social networks~\cite{gomez-rodriguez_inferring_2012,tang_social_2009}, our results further indicate that the social information can also help improve the prediction performance of image-based tasks. Among the three social factors, stable influence ($f_5$) has the most contribution. This phenomenon further indicates the users' emotion influence is not short-lived but last for some time, which verifies the rationality of the proposed factors' definitions.
\begin{table*}
  \centering
  \caption{Performance comparison between our model and SVM, and contribution of different factor functions by performance comparison of emotion prediction (\%). Model-$f_3$ represents removing temporal correlation factor, Model-$f_4$ represents removing social influence factor, Model-$f_5$ represents removing stable influence factor.}
  \label{tab:factorContr}
  \begin{tabular}{|m{1.4cm}|m{0.9cm}|m{0.9cm}|m{1.37cm}|m{1.37cm}|m{1.37cm}|m{0.9cm}|m{0.9cm}|m{1.37cm}|m{1.37cm}|m{1.37cm}|} \hline
    Emotional  & \multicolumn{5}{|c|}{Accuracy} & \multicolumn{5}{|c|}{F1-Measure} \\ \cline{2-11}
     categories& SVM & Model & Model-$f_3$ & Model-$f_4$ & Model-$f_5$ & SVM & Model & Model-$f_3$ & Model-$f_4$ & Model-$f_5$ \\\hline
     Happiness&63.74& 61.40 & 59.65 & 60.23 & 59.94 &66.67&  63.93&62.70&62.84& 62.47 \\\hline
     Surprise&72.80& 77.78 & 77.49 & 75.73 & 75.73 &34.04&38.71&38.40&34.65&34.65  \\\hline
     Anger&73.09& 77.19 & 76.02 & 75.73 & 75.73 &25.80&26.42&25.45&25.23& 25.23 \\\hline
     Disgust&78.94& 83.63 & 85.09 & 83.63 & 83.63 &28.00&33.33&35.44&33.33&33.33  \\\hline
     Fear&69.29& 73.98 & 72.81 & 72.81 & 72.51 &37.12&42.58&41.51&42.24&41.98  \\\hline
     Sadness&71.05& 76.02 & 73.98 & 74.85 & 74.85 &37.73&37.88&36.88&37.68&37.68  \\\hline
     \textbf{Average}&\textbf{71.49}&\textbf{75.00} &\textbf{74.17}&\textbf{73.83}&\textbf{73.73}&\textbf{38.23}&\textbf{40.48}&\textbf{40.06}&\textbf{39.33}&\textbf{39.22}\\\hline
  \end{tabular}
\end{table*}

\subsection{Case Study of Emotion Influence}
After objective evaluations on the propose model in predicting emotions of images, we use the total 4,725 users and their 1,254,640 images to infer the emotion influence.
%How to demonstrate the rationality of the predicted emotion influence?
Since it is very hard to find an objective way to evaluate the influence results, we would like to show an interesting case to demonstrate the effectiveness of the model.

\begin{figure}[htb]
  \centering{\includegraphics[width=0.5\textwidth]{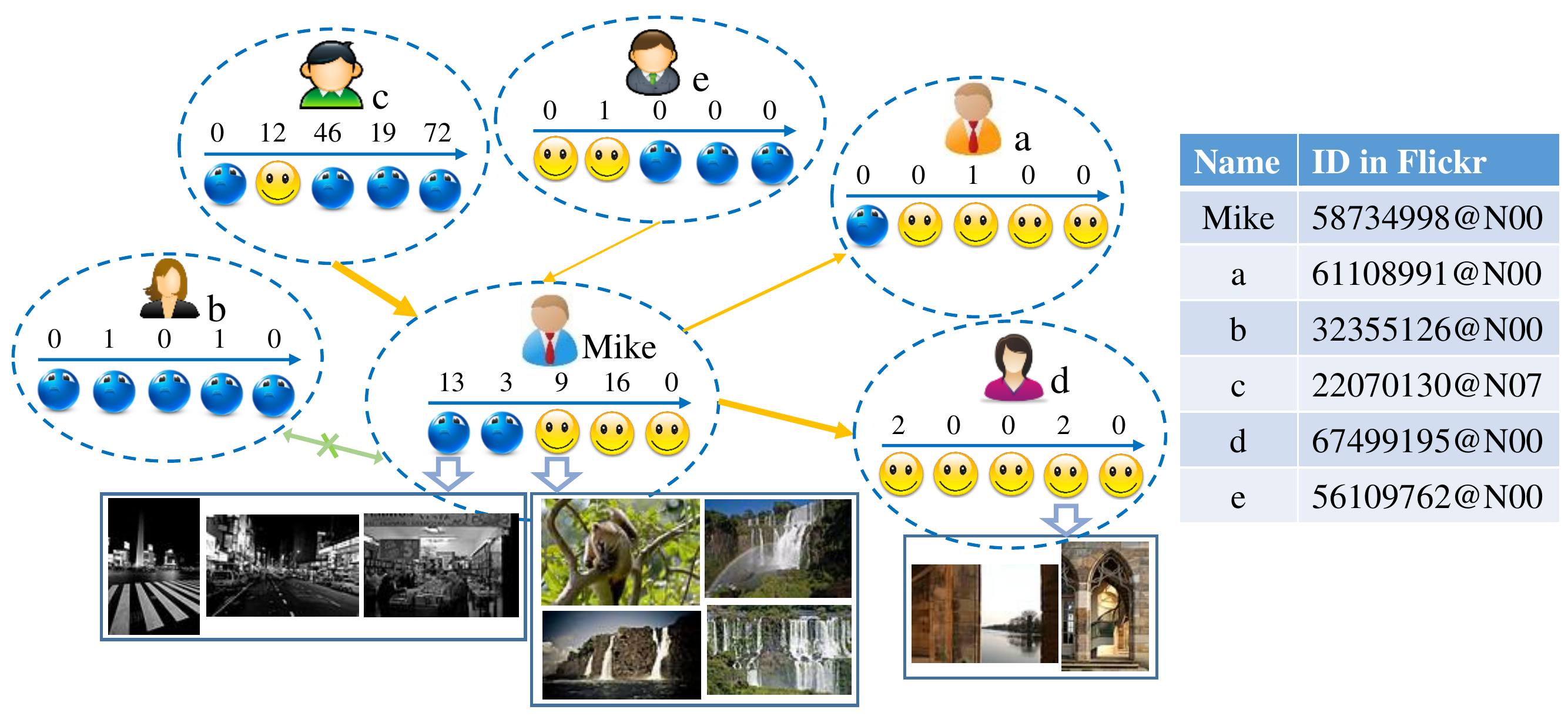}}
\caption{Case study: visualization of emotion influence between \textit{Mike} and his friends. For each user, it shows the numbers of uploaded images and the predicted emotions in the last five weeks. The arrow indicates the influence. The greater the impact is, the thicker the arrow is.}
\label{fig:caseStudy}
\end{figure}

Here we randomly chose a user from Flickr, called \textit{Mike} here (ID: 58734998@N00). During 12 weeks after January 26, 2006, he uploaded 119 images. \figref{fig:caseStudy} shows the visualization of some significant ``happy'' influence among his friends and him as an example. We can summarize the situation as follow: during that period, his mood used to be influenced by two friends, named as \textit{c} and \textit{e}, and he was more affected by \textit{c}. While he brought other two friends \textit{a} and \textit{d} the happy feelings, and he impacted \textit{d} more. Besides, there was no influence between \textit{Mike} and \textit{b}.

Let's further see this case in detail. During the 11th week, \textit{Mike} shared 16 images in total, which are all predicted to ``happy''. All these images are sightseeing such as waterfall or lovely animals with bright and vivid colors. He seemed to be in a great travel, certainly in a good mood. Later, one of his friends, called \textit{d} here, shared two images. Those are the only two images he shared during this week, which are about a beautiful building with also the bright colors. Our model predicted both images to ``happy'', and inferred the influence from \textit{Mike} to \textit{d}. That is, you are happy, I am happy. These results are reasonable and consistent with our observation, which indicates the effectiveness of our model.

\section{Conclusions}
\label{sec:conclusion}
In this paper, we study a novel problem of modeling emotion influence from images in social networks. We first examine the existence of the image-based emotion influence among friends. Then, we propose an emotion influence model to combine uploaded images, users' emotion and their influence.
We evaluate the proposed model on a large image dataset. The experimental results show that our model is much more effective on inferring emotions from social images than traditional SVM method. We also use a case study to further demonstrate that the prediction of the emotion influence is consistent with the reality.

Emotion is one of the most important user behaviors and presents a fundamental research direction.
%can influence people's actions
As the future work, it is intriguing to extend the peer influence study to group conformity behavior, such as how a user's emotion conforms to the community she/he belongs to. It is also interesting to further study
how the users' emotions correlate with their social actions.
There are many real applications of emotion analysis such as recommendation and social advertising. 

\balance
{\small
\bibliographystyle{ieee}
\bibliography{egbib}
}

\end{document}